# Comparison of hard x-ray production from various targets in air using a short pulse kHz laser with photon production from a high power multifilament laser beam from the same targets in air.

K.W.D.Ledingham*[1], S. S.Abuazoum[1], T.McCanny[1], J.J.Melone[1], K.Spohr[2], U.Schramm[3], S.D.Kraft[3], A.Wagner[3], A.Jochmann[3]

[1]SUPA, Dept of Physics, University of Strathclyde, Glasgow G40NG, UK
[2]SUPA, School of Engineering and Science, University of the West of Scotland, Paisley, PA12BE, UK
[3]Helmholtz Center Dresden-Rossendorf, 01328 Dresden, Germany.
*Corresponding author: k.ledingham@phys.strath.ac.uk

**Abstract:** Over the last few years there has been much interest in the production of hard X-rays from various targets using a kHz short pulse laser at intensities above $10^{14}$Wcm$^{-2}$ (A). Most of these studies have been carried out in vacuum and very many fewer studies have been carried out in air. Recently this lack has been partially addressed with the development of femtosecond laser micromachining. Another similar although apparently unconnected field (B) deals with the channelling of high power laser beam in filaments after passage through long distances in air. This has been largely driven by the construction of a mobile terawatt laser beam (Teramobile) for atmospheric studies. The laser beams in these two cases (A and B) have very different pulse energies (mJ against J) although the filaments in (B) have similar energies to (A) and are clamped at intensities less than $10^{14}$ Wcm$^{-2}$. This paper has been written to compare the production of hard X-rays in these two cases. The conclusion is interesting that a focused sub TW laser pulse in air reaches intensities sufficiently high that characteristic K and L X-rays are generated from a number of metal and non metal targets as well as a continuous bremsstrahlung spectrum. On the other hand the clamping of the multi-filaments in a 100 TW laser beam in air cannot generate hard X-rays especially when propagated over long distances.

**1 Introduction:**

The interaction of focussed terawatt and petawatt laser beams with solid and gaseous targets to produce multi MeV beams of electrons and X-rays has proved to be a very popular field of research in the last few years particularly with applications to nuclear physics and medicine in mind e.g. **(1-12).** It is important to emphasise that this work has normally been conducted under vacuum conditions with high pulse energies and usually with low repetition-rate lasers.

On the other hand recently laser based, high repetition rate, high power, compact X-ray sources in vacuum have been developed for a variety of applications particularly since access to third and fourth generation radiation synchrotron sources is and will continue to be limited **(13-17)**

Studies in air on the other hand are generally lacking and the physics and the outcome involved in transmitting high power laser beams through air to hit targets is very different from through vacuum. There are two principle applications for directing high power lasers through air. Firstly **(18-22)** a number of groups are interested in X-Ray production in air for femtosecond laser micromachining and other surface science applications such as X-ray fluorescence studies for analysis. The second application is beam filamentation and white light continuum production and divergence as a high power short pulse laser propagates through air or transparent condensed matter. Ultra-short (fs) high power laser pulses may exceed the threshold intensity for non-linear self-focusing in air resulting in the propagation of intense laser beams for many meters and even kilometres in the atmosphere **(23)**. The applications for this laser technology are many e.g. remote sensing, lidar, lightning control **(24)** and laser induced water condensation in air and have lead to the construction of the first mobile terawatt laser (Teramobile) for atmospheric studies. More recently it has been suggested that the associated intense electric fields might generate secondary radiation which could disrupt electronic devices resulting in sensor damage and even the construction of a possible counter mine technology. The theory associated with the passage of intense lasers through air and the applications mentioned above has lead to the publication of many papers only some of which are referenced here **(25-38)**

This paper has been written to compare and contrast X-ray production when a single unfilamented laser beams interacts with various targets in air using a short pulse kHz laser with X-ray production when a high power multifilament laser beam interacts with the same targets in air.

**2. Experimental**

This section will be presented in two parts a) dealing with x-ray production in air from various targets using a high power, high repetition rate, fs laser in a single focussed beam; b) dealing with photon production from a high power, multiterawatt multifilamented beam after passage through ten metres of air. In the conclusions the results from the two sections will be compared and contrasted and suggestions proposed for future research.

*2.1 X-ray production in air from various targets using a high power, high repetition rate, fs laser in a single focussed beam.*

The experiments were performed using a Thales Alpha 1000 laser producing pulses of 100 fs duration at a wavelength of 800 nm with a 1 kHz repetition rate. The laser was focussed in air with a 10 cm lens to a spot size of 50-100 microns resulting in laser intensities (I) of between $10^{15-16}$ Wcm$^{-2}$ (determined from the pulse energy and the laser spot size) and directed normally to various targets of about 1 mm thickness. The X-ray production was measured using an AmpTec XR-100CR detector 6 mm in diameter and 0.5 mm thick with a 0.5 mil Be window operated in single photon detection mode. This is a Si PIN photodiode thermoelectrically cooled at approx -55° with coupled preamplifier and cooler system. The resolution of the detector was about 160 eV, easily able to resolve the $K_{\alpha,\beta}$ flourescent x-rays of low Z atoms. The detector was placed in reflective mode at 45° to the normal and about 10 cm from the focal spot. The target was continuously translated to prevent the beam spot from digging a pit which reduced the x-ray production. The target assembly was shielded by lead bricks to eliminate any radiation hazard. The spectra were obtained using a computer based multichannel analyser and the spectrum obtained from an iron target is shown in Fig. 1. It is very difficult to suppress pileup when conventional nuclear detectors with microsecond shaping electronics are used in a short pulse (fs) laser environment and the inset to Fig. 1. shows pile up artifacts up to the detection of 5 simultaneous iron $K_\alpha$ x-rays when the laser intensity was high. The K x-ray yield has been shown to have approximately an $I^3$ dependence at laser intensities between $1-3 \times 10^{15}$ Wcm$^{-2}$ **(19).** The fluorescent x-rays sit on a continuous background of hard photons produced by bremsstrahlung from energetic electrons in the plasma entering the target. A review in the production of X-rays both line and broadband emissions was made in 2001 **(39).** The spectrum in Fig 2 shows L x-rays from a cadmium target. These spectra were averaged over typically many thousand shots. Similar spectra have been obtained by other authors **(18 -20).**

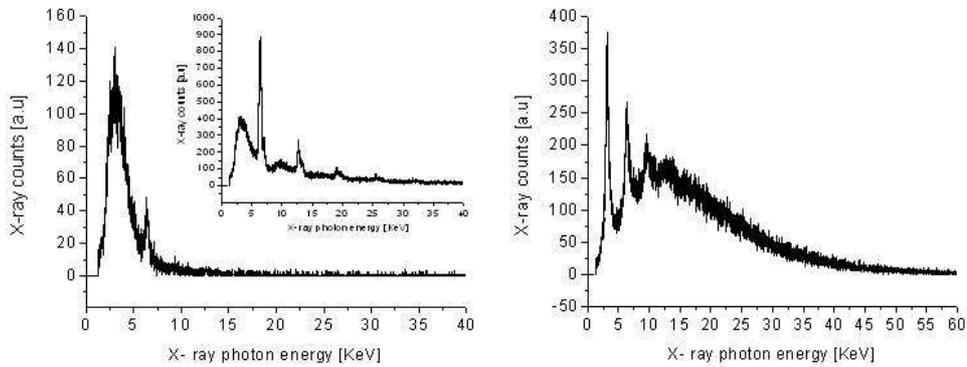

Fig. 1. X-ray spectra. Left : the iron K x-ray sitting on a laser induced bremsstrahlung spectrum. The inset shows a number of pileup artifacts when the laser intensity is increased. Right: Laser induced x-ray spectrum from a Cd target showing characteristic L peaks on a continuous bremmstrahlung spectrum

In a seminal paper describing the excitation of x-rays by electrons accelerated in air in the wake field of a laser field the authors show a spectrum of continuous hard x-rays and $K_\alpha$ x-rays as a function of laser intensity between $10^{14-15}$Wcm$^{-2}$. This is shown with the authors permission in Fig 2 **(25).**

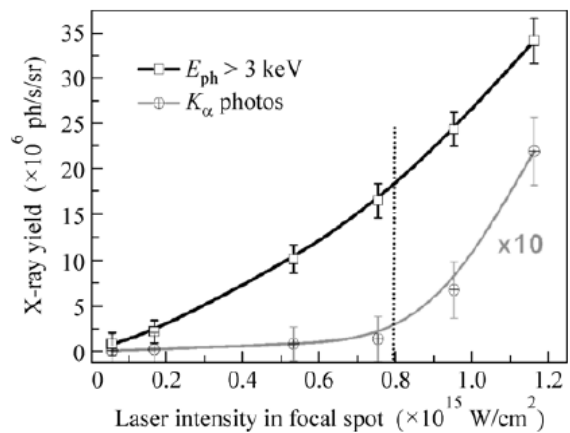

Fig. 2. Yield of x-rays from a copper target in air as a function of laser intensity. The grey line is the intensity of the Cu $K_\alpha$ line and the solid line is the total x-ray intensity >3keV which is essentially a function of the bremsstrahling spectrum. Reproduced by kind permission of the authors **(25)**.

*2.2 Photon production from a high power, multiterawatt multifilamented beam with various targets after passage through ten metres of air.*

In recent years the filamentation of ultrashort (fs) intense laser pulses has generated much theoretical interest in the laser community when the laser is passed through transparent media like liquids and gases. Self channeling of laser energy in a filament over many tens or hundreds of meters is a complicated process essentially as a result of a dynamic equilibrium between Kerr self focusing which causes photoionisation and then defocusing in the induced laser plasma when the laser power is greater than about 3 GW **(36,37)**. The filaments are distributed in apparently a random but steady fashion which are shown in Figs 3. and 5. Depending on the intensity of the laser many hundreds of filaments can be distributed over the dimensions of an unfocused multi terawatt beam. The diameter of the filaments is about 10-100μ in diameter and all have similar diameters **(28)**. It has been noted by a number of the authors that the intensity in a filament is clamped between 5-7x$10^{13}$Wcm$^{-2}$ **(26,27)**. Moreover it has been measured **(34)** that the clamped intensity for a specific gas hardly changes when the laser intensity increased by a factor as much as a hundred and even when two filamented laser beams are added coherently the clamped intensity hardly changes **(40)**. Although filamentation is normally associated with the passage of intense lasers through transparent materials as mentioned above, filamentation has also been observed at the rear of targets in vacuum caused by Weibel-like instabilities in laser produced electrons beams (**12**).

The experiment was performed on the 100 TW femtosecond Ti:sapphire laser (DRACO) at Helmholtz-Zentrum Dresden-Rossendorf at a wavelength of 800 nm. In the present experiment the unfocused laser pulse energy was about 2.7 J with a pulse length of 60fs, a beam diameter of 80 mm and a pulse repetition rate of 10 Hz. The beam was reflected using two mirrors into the air and the induced filamentation phenomena was observed after passage through about 10 m of air to hit targets of different materials. The target was surrounded by concrete blocks for safety reasons. The experimental arrangement is shown diagramatically in Fig.3. and in real life in Fig. 4. The filaments on the target and the filaments through the air are shown in Fig 5. The filaments in air could be seen when the dust ablated from a Pb target scattered light from the laser beam.

The experimental arrangement to measure the radiation detected after reflection from different targets was analysed by two detectors placed at 45° to the direction of the laser beam. This is shown in diagrammatic form in Fig. 3. and in real life in Fig 4.

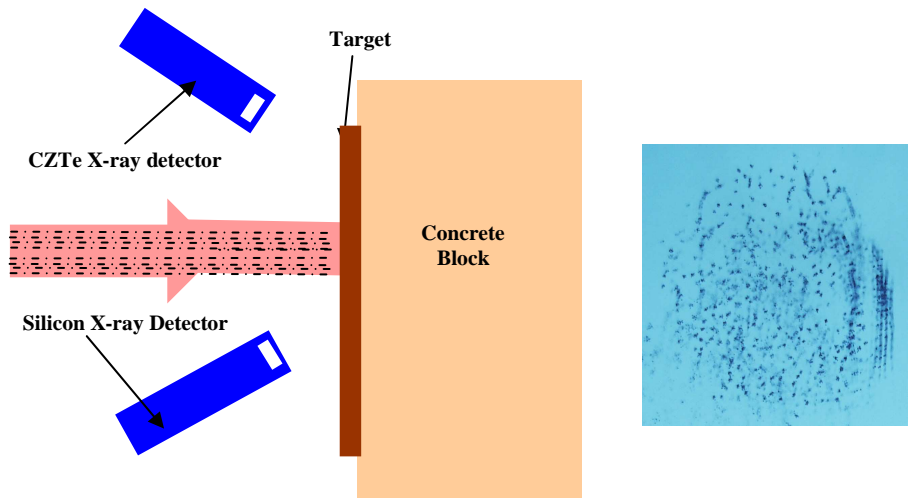

Fig. 3. This figure shows a diagram of the detector arrangement to detect any x-rays which are emitted after the interaction of the multi-filamented laser beam with a number of different targets both metal and non metal (concrete). The target area was surrounded with concrete blocks as a safety precaution. The right hand blue diagram shows the interaction of the filaments with burn paper after a single shot.

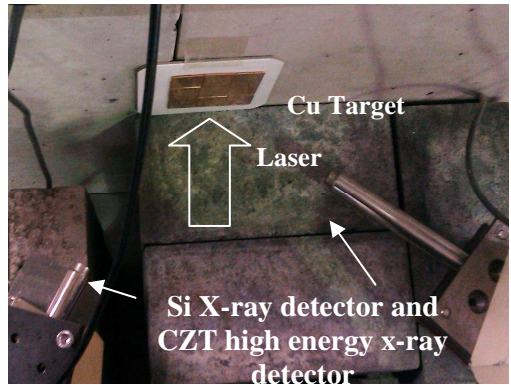

Fig. 4. This photograph shows the interaction of the filamented laser beam with a copper target. The detectors (Si and CZT ) are placed at 45º to the laser beam and at 16 and 12 cm respectively. The CZT detector can has a much greater efficiency for the detection of higher energies x-rays than the Si PIN detector. The assembly is surrounded by concrete blocks for radiation safety.

One was a Si PIN detector and the other a higher efficiency CZT detector. The radiation monitors were operated in single pulse mode and the spectra obtained over thousands of laser pulses were analysed by a multichannel pulse height analyser. The presence of filamentation is demonstrated in Fig. 5.

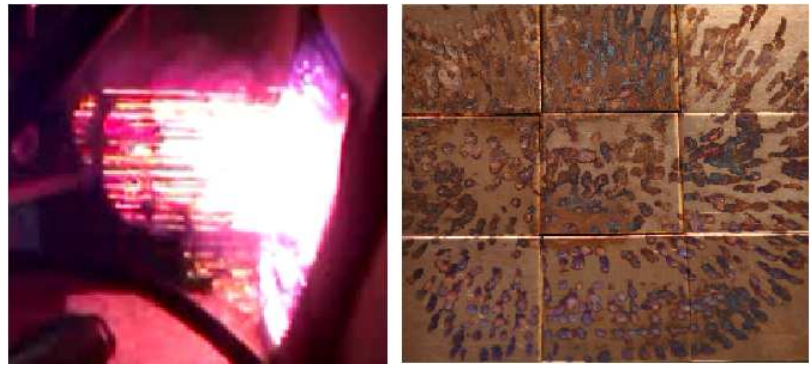

Fig.5. This figure shows filaments in air interacting with a Pb target (left). The ablated Pb dust particles reflect the light from the filaments making them visible in ambient daylight. The photograph on the right provides a permanent record the interaction of the filaments with the copper target after many hundreds of shots.

The spectra obtained from lead, aluminium and concrete targets are shown in Fig. 6. The calibration of these spectra was carried out using the L x-rays from a $^{241}$Am source which has peaks from 9-27 keV. If the filamented intensities were sufficient to cause the production of hard x-rays then the fluorescent K and L x-rays would be visible in these spectra. The intensity of these spectra was greatly reduced when thin copper absorbers were placed in front of the detectors and also no radiation was transmitted through the metal targets. It is known that the filamented intensity is sufficient to generate the characteristic x-rays of oxygen and nitrogen (see Fig.7B) and hence a count at e.g. 20 keV would not necessarily indicate the detection of a 20 keV photon but could arise from the simultaneous absorption of a number of lower energy photons as was shown in Fig.1. The absorption of these events would of course be the characteristic of the absorption of the much lower energy photons. This might indicate that the spectra obtained were some sort of low energy pileup artefacts.

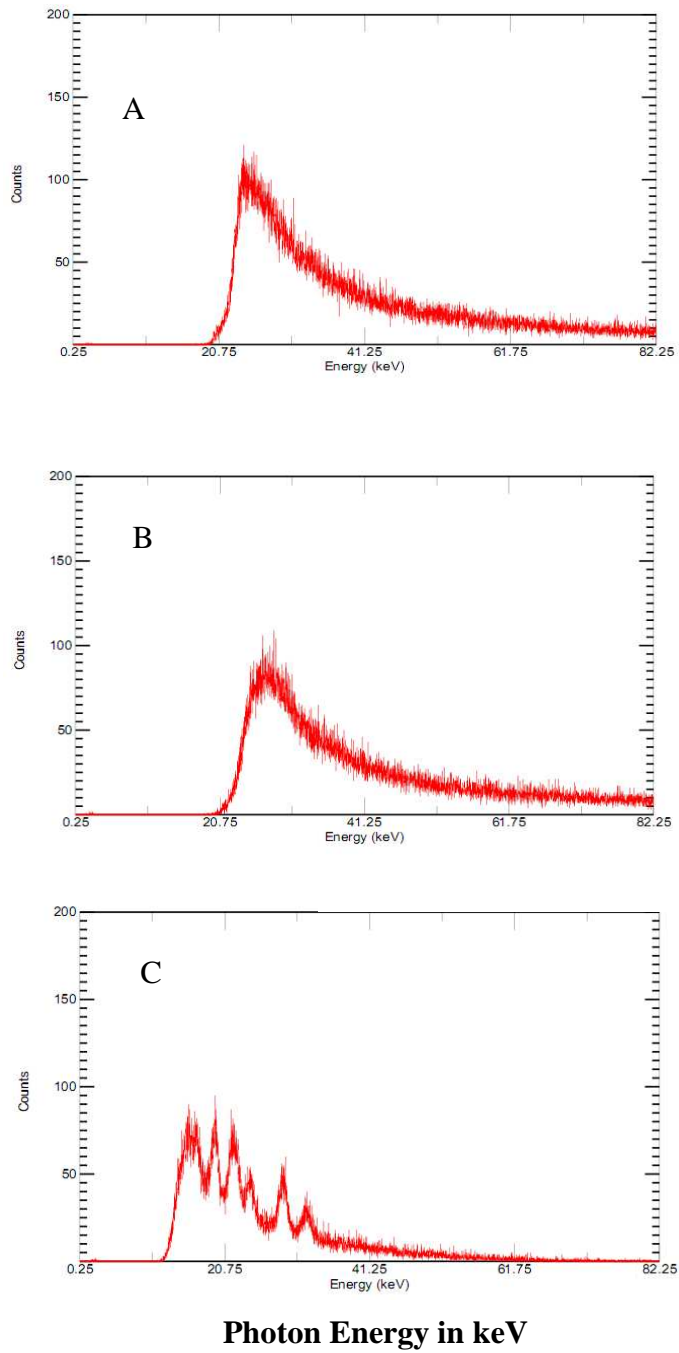

**Photon Energy in keV**

Fig. 6. The three spectra above show the radiation detected by the CZT detector from A) lead, B) aluminium and C) concrete in reflection mode. The spectra from the metal samples are very similar in intensity when normalised for time without showing any Z dependence or characteristic peaks. The concrete spectrum is peaked (characteristic of many spectra with different types of concrete )which is almost certainly caused by the non conducting form of target and may well be discharge.

## 3. Conclusions

When a single focused beam from a short pulse (fs) high repetition rate Ti: sapphire laser beam interacts with different metal and non-metals in air at intensities above $10^{14} Wcm^{-2}$ both characteristic X-rays and broad band hard X-rays are detected. The intensities of the X-rays increase strongly with laser intensity. It has been shown that hard x-rays production falls to zero as the intensity is reduced below $10^{14} Wcm^{-2}$ (**25**). In this case it is clear that the laser has not travelled sufficiently far in air to cause filamentation to take place.

On the other hand no characteristic X-ray production is observed when a heavily filamented unfocused laser beam interacts with different metal and non metal targets. This can be understood since the intensity of the filamented beam is clamped at between $5-7 \times 10^{13} Wcm^{-2}$ (**26,27**) where it has been shown that no hard X-rays are formed. Even in tight focusing conditions, there is no X-ray production because the clamped intensity is not enhanced by much even when the pulse energy is increased by a factor of 100 (**34**). The decrease of the beam diameter did however result in the detected fluorescent from the air become more stable.

The characteristic white light continuum generated in air by a high power laser is shown in Fig. 7. and does show that line spectra from the gases in air are generated but these spectra are in the UV. This of course makes it a very important procedure in the field of atmospheric research for detecting trace gases (pollutants) at long distances and also laser triggered lightning and rain (**37,38,41**).

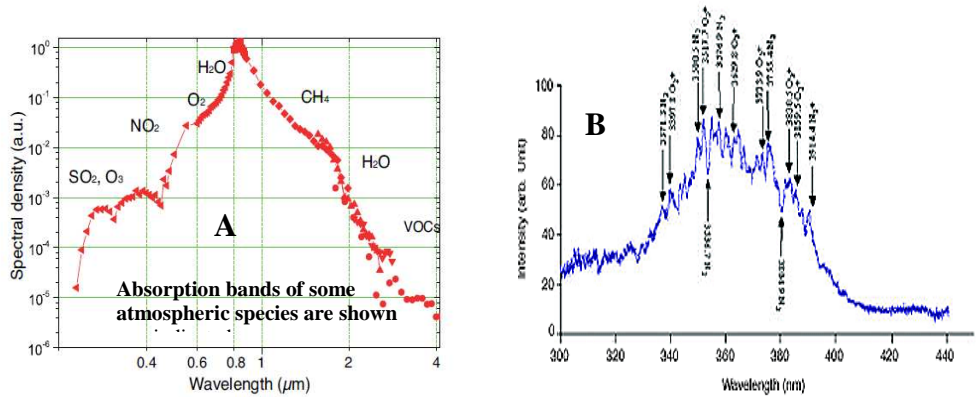

Fig. 7 A) White light continuum spectrum of a 70fs laser beam, 800nm, 3 TW, after propagation over a distance of 10 m in air. (Reproduced by kind permission of the authors (**24**)). B) A linear broadband emission spectrum in the UV from a 100fs, 0.3TW laser after passage through 7 m in air. The line spectra have been identified as optical transitions in neutral and ion species in nitrogen and oxygen (Reproduced by kind permission of the authors (**35**))

Theoretically the lack of x-rays is what one might expect. At the clamped intensity of $\sim 10^{14} Wcm^{-2}$ it is not surprising that no x-rays are detected since the relativistic ponderomotive potential ($I_p$) (**42**)

$$I_p = 0.511(\sqrt{1+I\lambda^2/1.37 \times 10^{18}} -1) \text{ MeV}$$
$$= 0.511\{\sqrt{1+10^{14} \times 0.64 /1.37 \times 10^{18}} -1)$$
$$= 12 \text{ eV for 800 nm or 120ev for } 10^{15} Wcm^{-2}$$

Thus we might have bremsstrahlung up to these energies and hence it would be difficult to excite characteristic x-rays in targets. Now this calculation is in vacuum and so the energy would be somewhat reduced in atmospheric conditions. If hard x-rays were generated by higher harmonics which would stretch up to ~ $3I_p$ = ~36 eV for 0.8μ and $10^{14}$Wcm$^{-2}$ or 360ev for $10^{15}$Wcm$^{-2}$. Thus from these simple calculations it is difficult to see how hard x-ray energies up to few keV would be generated in a filament. Curiously it has been reported that hard x-rays have been generated in air (**43**) by focussing a 25ps laser on a copper target at an intensity of $3\times10^{13}$ Wcm$^{-2}$ which is slightly lower than the clamped intensity of a filament. The authors are uncertain of the acceleration mechanism although they suggest that resonant absorption during the laser–matter interaction might be an explanation for the production of hot electrons. This finding has yet to be verified.

It was described earlier that recently it has been suggested that the associated intense electric fields associated with lasers in air might generate secondary radiation which could disrupt electronic devices resulting in sensor damage and even the construction of a possible counter mine technology (**36**). Certainly there was no evidence of secondary hard X-ray emission from the filamentation in air. Hence the only possibility of generating sufficiently intense secondary radiation to disrupt electronic devices would be to control the multi-filamentation in air by beam astigmatism (**44**) or if it were possible to coherently add the many hundreds of filaments, a procedure which has to date proved elusive.

In summation multifilamented laser beams can deliver over considerable distances high (~$10^{14}$Wcm$^{-2}$) but clamped intensities. If this is sufficient for any proposed application then this is a considerable advantage. For example the clamped intensity allows laser induced breakdown spectroscopy to operate for the remote detection of mines and other explosive materials (**45**). If however higher intensities are required remotely then present knowledge does not allow one to defeat the clamping.

**Acknowledgements**

It is a pleasure to acknowledge many, useful and informative discussions with SeeLeang Chin, Jerome Kasparian, Kamil Stelmaszczyk (Teramobile team) and Robert Deas (DSTL).

**References and Links**

1. K.W.D.Ledingham and W. Galster, "*Laser –driven particle and photon beams and some applications*" New J.Phys., **12,** 045005- 045071 (2010)
2. K.W.D.Ledingham, I.Spencer, T.McCanny, R.P.Singhal, M.I.K.Santala, E.Clark, I.Watts, F.N.Beg, M.Zepf, K.Krushelnick, M.Tatarakis, A.E.Dangor, P.A.Norreys, R.Allott, D.Neely, R.J.Clark, A.C.Machacek, J.S.Wark, A.J.Creswell, D.C.W.Sannderson , J.Magill " *Photonuclear Physics when a Multiterawatt Laser Pulse Interacts with Solid Targets*" Phys.Rev.Lett **84,** 899-902 (2000)
3. T.E.Cowan, A.W.Hunt, T.W.Phillips, S.C.Wilks, M.D.Perry, C.Brown, W.Fountain, S.Hatchett, J.Johnson, M.H.Key, T.Parnell, D.M.Pennington, R.A.Snavely and Y Takahashi, "*Photonuclear Fission from High Energy Electrons from Ultraintense Laser-Solid Interactions*" Phys Rev.Lett., **84,** 903-2006 (2000)
4. J. Faure, Y.Glinic, A.Pukhov, S.Kiselev, S.Gordienko, E.Lefebre, J-P.Rousseau, F.Burgy ,V.Malka "*A laser-plasma accelerator producing monoenergetic electrons beams*" Nature **431** 541-544(2004)
5. C.G.R.Geddes, C.S.Toth, J.Van Tilborg, E.Esaray, C.B.Schroeder, D.Bruhwiler, C.Nieter, J.Cary, W.P.Leemans"*High–quality electron beams from a laser wakefield accelerator*" Nature **431** 538-541 (2004)
6. S.P.D.Mangles, C.D.Murphy, Z.Najmudin, G.R.Thomas, J.L.Collier, A.E.Dangor, E.J.Divall, P.S.Foster, J.G.Gallacher, C.J.Hooker, D.A.Jaroszynski, A.J.Langley, W.B.Mori, P.A.Norreys, F.S.Tsung, B.R.Walton, K.Krushelnick "*Monoenergetic beams of relativistic electrons from intense laser-plasma Interactions*" Nature **431** 535-538 (2004)
7. V.Malka , J.Faure, Y.A.Gauduel, E.Lefebre, A.Rousse ,K.T.Phuoc "*Principles and Applications of compact laser-plasma accelerators*" Nature Physics **4**, 447-453 (2008)
8. H. Schwoerer , P.Gibbon, S.Dusterer, R.Behrens, C.Ziener, C.Reich, R.Sauerbrey "*MeV xrays and Photoneutrons from Femtosecond Laser–Produced Plasmas*" Phys.Rev.Lett., **86,** 2317-2320(2001)
9. J.D.Kmetec , C.L.Gordon III, J.J.Macklin, B.E.Lemoff, G.S.Brown, S.E.Harris "*MeV X-Ray Generation with a Femtosecond Laser*" Phys.Rev.Lett. **68,** 1527- 1530 (1992)
10. K.M.Spohr, M.Shaw, W.Galster, K.W.D.Ledingham, L.Robson, J.M.Yang, P.McKenna, T.McCanny J.J.Melone, K-U Amthor, F.Ewald, B.Liesfeld, H.Schwoerer, R.Sauerbrey, "*Study of Photo-Proton Reactions driven by bremsstrahlung radiation of high-intensity laser generated electrons*" New J.Phys. **10,** 043037 (2008)
11. C.Gahn, G.Pretzler, A.Saemann, G.D. Tsakiris, K.J.Witte, D.Gassmann, T.Schatz, U.Schramm, P.Thiroff, D.Habs " *MeV γ-ray yield from solid targets irradiated with fs-laser pulses*" Appl.Phys.Lett.

**73,** 3662-3664 (1998)
12. M.S.Wei, F.N.Beg, E.L.Clark, A.E.Dangor, R.G.Evans, A.Gopal, K.W.D.Ledingham, P.McKenna, P.A.Norreys, M.Tatarakis, M.Zepf, K.Krushelnick, *"Observations of the filamentation of high-intensity laser produced electron beams"* Phys.Rev.E **70,** 065412-056416 (2004)
13. Y.Jiang, T.Lee, C.G.Rose-Petruck, *"Generation of ultrashort hard x-ray pulses with tabletop laser systems at a 2-kHz repetition rate"* J.Opt.Soc.Am.B **20,** 229-236 (2003)
14. F.Gobet, F.Hannachi, M.M.Aleonard, J.F.Chemin, G.Claverie, M.Gerbaux, G.Malka, J.N.Scheurer, M.Taisien, F.Blasco, D.Descamps, F.Dorchies, R.Fedosejevs, C.Fourment, S.Petit, V.Meot, P.Morel, S.Hanvey, L.Robson, B.Liesfeld *"Absolute energy distribution of hard x-rays produced in the interaction of a kilohertz femtosecond laser with tantalum targets"* Rev.Sci.Instrum.,**77,** 093302.(2006)
15. A.Sjogren, M.Harbst, C.-G.Wahlstrom, S.Svanberg, *"High–repetition-rate, hard x-ray radiation from a laser-produced plasma: Photon yield and application considerations"* Rev.Sci.Instrum.**74,** 2300-2311 (2003)
16. A.Thoss, M.Richardson, G.Korn, M.Faubel, H.Stiel, U.Vogt, T.Elsaesser *"Kilohertz sources of hard x-rays and fast ions with femtosecond laser plasmas"* J.Opt.Soc.Am B **20,** 224-228 (2003)
17. J.Kutzner, H.Witte, M.Silies, T.Haarlammert, J.Huve, G.Tsilimis, I.Uschmann, E.Forster, H.Zacharias *"Laser Based, high repetition rate, ultrafast X-ray source"* Surface and Interface Analysis **38,**1083-1089(2006)
18. K.Hatanaka, T.Miura, H.Fukumura, *" Ultrafast x-ray pulse generation by focusing femtosecond infrared laser pulses onto aqueous solutions of alkali metal chloride"* Appl. Phys.Lett., **80,** 3925 (2002)
19. J.Thogersen, A.Borowiec, H.K.Haugen, E.E.McNeill, I.M.Stronach *"X-ray emission from femtosecond laser micromachining"* Appl.Phys.A **73,** 361-363(2001)
20. J.Hatanaka, K.Yomogihata, H.Ono, H.Fukumura *" Femtosecond laser induced X-ray emission from metal alloys polymers and color filters"* Appl.Surface Sci., **247,** 232 (2005)
21. P.Scully, D. Jones, D.A. Jaroszynski, *" Femtosecond laser irradiation of polymethylmethacrylate for Refractive index gratings"* J.Opt.A: Pure Appl.Opt **5,** S92-S96 (2003)
22. V.M.Gordienko, M.S.Dzhidzhoev, I.A.Zhvaniya, I.A.Makarov *" Increase in the yield of X-ray photons upon two-laser excitation of a solid target in air"* Quantum Electronics **37,** 599-600 (2007)
23. Y. Petit, S.Henin, W.M.Nakaema, P.Bejot, A.Jochman, S.D.Kraft, S.Bock, U.Schramm, K.Stelmaszczyk, P.Rohwetter, J.Kasparian, R.Sauerbrey, L.Woste, J-P. Wolfe *"1-J white-light continuum from 100 TW laser pulses"* Phys.Rev.A. **83,** 013805 (2011)
24. J.Kasparian J-P.Wolf *"Physics and applications of atmospheric nonlinear optics and filamentation"* Optics Express, **16,** 466-493(2008)
25. A.Ya.Faenov, S.A.Pikuz Jr, A.G.Zidkov, I.Yu.Skobelev, P.S.Komarov, O.V.Chefonov, S.V.Gasilov, A.V.Ovchinnikov, *"Ezcitation of X-Rays by Electrons Accelerated in Air in the Wake Wave of a Laser Pulse"* JETP Letters, **92,** 375 (2010)
26. J.Kasparian, R.Sauerbrey, S.L.Chin *" The critical laser intensity of self-guided filaments in air"* Applied Physics B Lasers and Optics, **71,** 877- 879 (2000)
27. A..Braun, G.Korn, X.Liu, Du.J.Squier, G.Mourou *"Self-channeling of high-peak-power femtosecond laser pulses in air"* Optics Letters **20**, 73-75 (1995)
28. A.Brodeur, F.A. Ilkov, S.L.Chin, *"Beam Filamentation and the White Light Continuum Divergence"* Optics Communications, **129,** 193(1996)
29. J.Kasparian, R.Sauerbrey, D.Mondelain, S.Niedermeier, J.Yu, J.-P. Wolf, Y.-B. Andre, M.Franco, B.Prade, S.Tzortzakis, A.Mysyrowicz, M.Rodriguez, H.Wille, L.Woste, *"Infrared extension of the supercontiuum generated by femtosecond terwatt laser pulses propagating in the atmosphere"* Optics Letters , **25,** 1397-1399 (2000)
30. S.A.Hosseini, J.Yu, Q.Luo, S.L.Chin *" Multi-parameter characterisation of the longitudinal plasma profile of a filament: a comparative study"* Applied Physics B Lasers and Optics, **79,** 519-523 (2004)
31. Q.Luo, S.A.Hosseini, W.Liu, J.-P. Gravel, O.G.Kosareva, N.A.Panov, N.Akozbek,, V.P.Kandidov, G.Roy, S.L.Chin *" Effect of beam diameter on the propagation of intense femtosecond laser pulses"* Applied Physics B Lasers and Optics, **80,** 35-38 (2005)
32. L. Roso, J.San Roman, I.J.Sola, C.Ruiz, V.Collados, J.A.Perez, C.Mendez, J.R.Vasquez de Aldana, I Arias, L.Plaja, *Peopagation of terwatt laser pulses in the air"* Applied Physics A Materials Science and Processing, **92,** 865-871 (2008)
33. J.Zhang, X.Lu, Y.Y.Ma T.T.Xi, Y.T.Li, Z.M.Sheng, L.M.Chen, J.L.Ma, Q.L.Dong, Z.H.Wang, Z.Y.Wei*," Long Distance Femtosecond Laser Filamaments in Air"* Laser Physics, **19,** 1769 -1775 (2009)
34. O.G.Kosareva, W.Liu, N.A.Panov, J.Bernhardt, Z.Ji, M.Sharifi, R.Li, Z.Xu, J.Liu, Z.Wang, J.Jui, X.Lu, Y.Jiang Y.Leng, X.Liang, V.P.Kandidov, S.L.Chin *"Can We Reach very High Intensity in Air with Femtosecond Pw Laser Pulses* Laser Physics, **19,** 1776 (2009)
35. V.Y.Federov, O.V.Tverskoy,V.P. Kandidov, *"Transport of high fluence energy by femtosecond filament in air"* Applied Physics B Lasers and Optics, **99,** 299-306 (2010)
36. A.C.Ting, D.F.Gordon, R.F.Hibbard, J.R.Penano, P.Sprangle, C.K.Manka *" Filamentation and Propogation of Ultra-short, Intense Laser Pulses in Air"* http://www.nrl.navy.mil/research/nrl-review/2003/featured research/ting/
37. H. Wille, M.Rodriguez, J,Kasparian, D.Mondelain, A.Mysyrowicz, R.Saueerbrey, J.P.Wolf, L.Woste2002, *"Teramobile: A Mobile femtosecond laser and detection system"* Eur.Phys.J.AP **20,** 183-190 (2002)
38. P.Rohwetter, J.Kasparian, K.Stelmaszczyk, Z.Hao, S.Henin, N.Lascoux, W.Nakaema, Y.Petit, M.Queisser, R.Salame, E.Salmon, Ludger Woste, J-P. Wolf *" Laser Induced Water Condensation in Air"* Nature Photonics, **4,** 451-456 (2010)

39. J.-C. Kieffer, F.Dorchies, P.Forget, P.Gallant, Z.Jiang, H.Pepin, O.Peyrusse,, C.Toth, A.Cavalleri, J.Squier,K.Wilson "*Femtosecond Thermal X-ray Pulses from Hot Solid Density Plasmas*" Laser Physics **11,** 1201-1204 (2001)
40. S.Xu, Y.Zheng,Y. Liu, W.Liu, "*Intensity Clamping during Dual-Beam Interference*" Laser Physics **20,** 1968-1972 (2010)
41. K.Stelmaszczyk, P.Rohwetter, G.Mejean, J.Yu, E.Salmon, J.Kasparian, L.Woste *" Long-distance remote laser induced breakdown spectroscopy using filamentation in air"* Appl.Phys.Lett **85**, 3977-3979 (2004)
42. S.C.Wilks, W.L. Kruer , *"Absorption of Ultrashort, Ultra-Intense Laser Light by Solids and Overdense Plasmas*  IEEE, J.Quantum Electron, **33,** 1954-1967 (1997)
43. H.Nagao, Y.Hironaka, K.G.Nakamura, K.Kondo, *"Hard X-ray Emission from a Copper Target by focussing a Picosecond Laser Beam at $3x10^{13} Wcm^{-2}$"* Jap.J.Appl.Phys. **43,**1207-1208 (2004)
44. G.Fibich, S.Eisenmann, B.Ilan, A,.Zigler "*Control of Multiple Filamentation in Air"*  Optics Letters, **29,** 1772-1774, (2004)
45. W.Schade, C.Bohling, K.Hohmann, D.Scheel. "*Laser-induced plasma spectroscopy for mine detection and verification"* Laser and Particle Beams **24,** 241-247 (2006)